\documentclass[preprint,12pt]{aastex}

\shorttitle{\emph{APEX} CO (9--8) Mapping of NGC 6334 I}
\shortauthors{Qiu et al.}
\begin{document}

\title{APEX CO (9--8) Mapping of an Extremely High-Velocity and Jet-like Outflow in a High-Mass Star-Forming Region}

\author{Keping Qiu\altaffilmark{1,2}, Friedrich Wyrowski\altaffilmark{1}, Karl M. Menten\altaffilmark{1}, Rolf G\"{u}sten\altaffilmark{1},
Silvia Leurini\altaffilmark{1}, and Christian Leinz\altaffilmark{1}}\email{kqiu@mpifr-bonn.mpg.de}
\altaffiltext{1}{Max-Planck-Institut f\"{u}r Radioastronomie, Auf dem H\"{u}gel 69, 53121 Bonn, Germany}
\altaffiltext{2}{Key Laboratory of Modern Astronomy and Astrophysics (Nanjing University), Ministry of Education, Nanjing 210093, China}

\begin{abstract}
Atacama Pathfinder Experiment (APEX) mapping observations in CO (9--8) and (4--3) toward a high-mass star-forming region, NGC 6334 I,
are presented. The CO (9--8) map has a $6.\!''4$ resolution, revealing a $\sim$0.5 pc, jet-like, and bipolar outflow. This is the first
map of a molecular outflow in a THz line. The CO (9--8) and (4--3) lines arising from the outflow lobes both show extremely high-velocity
line wings, and their ratios indicate a gas temperature greater than 100 K and a density higher than $10^4$ cm$^{-3}$. The spatial-velocity
structure of the CO (9--8) data is typical of a bow-shock-driven flow, which is consistent with the association between the bipolar outflow
and the infrared bow-shaped tips. In short, the observations unveil a highly-excited and collimated component in a bipolar outflow that is
powered by a high-mass protostar, and provide insights into the driving mechanism of the outflow. Meanwhile, the observations demonstrate
that high-quality mapping observations can be performed with the new THz receiver on APEX.

\end{abstract}


\section{Introduction} \label{intro}
CO low-$J$ transitions have been widely observed in studies of molecular clouds and bipolar outflows. These lines, e.g., CO (1--0) and (2--1),
are easily excited at very low temperatures and mostly trace relatively cold molecular gas. Observations of high-$J$ CO lines, in particular
those falling in the THz frequency regime ($J_{\rm up}\geq9$), are very rare. These highly excited lines can be observed with spaceborne
observatories, e.g., Infrared Space Observatory \citep{Giannini01} and most recently, \emph{Herschel} Space Telescope
\citep{vKempen10,Yildiz10}. On the other hand, ground-based THz observations, which can provide better angular resolutions down to a few arc
seconds, are challenging even on extremely dry sites. To date only a very few ground telescopes have successfully obtained astronomical THz CO
spectra \citep[e.g.,][]{Kawamura02,Wiedner06}. However, all these observations were carried out targeting one or a few points in an observed
region, and had coarse spectral and/or spatial resolutions. Marrone et al. (2004) presented a CO (9--8) map of the OMC-1 region, but with a
poor spatial resolution of $84''$. Therefore, morphological and kinematical information of the gas probed by THz CO lines is missing.

As part of the science demonstration observations of the new THz receiver on APEX\footnote[3]{APEX is a collaboration between the
Max-Planck-Institut f\"{u}r Radioastronomie, the European Southern Observatory, and the Onsala Space Observatory}, we mapped the CO (9--8)
(1.037 THz) emission in NGC 6334 I, the brightest far-IR source in a relatively nearby \citep[1.7 kpc,][]{Neckel78} molecular
cloud/H{\scriptsize II} region complex \citep{McBreen79}. NGC 6334 I has been extensively studied over the last decade
\citep[e.g.,][and references therein]{Hunter06,Beuther07,Beuther08,Emprechtinger10,vdWiel10}. It contains an ultracompact (UC) HII region
first detected in 6 cm continuum \citep[labeled ``F'' in][]{Rodriguez82} and appears as a ``proto-Trapezium'' system in 1.3 mm continuum
\citep{Hunter06}. A bipolar molecular outflow, which is most relevant to this work, has been observed with single-dish telescopes in low- to
mid-$J$ CO lines and in SiO and CS lines \citep{Bachiller90,McCutcheon00,Ridge01,Leurini06}. Here the CO (9--8) line was observed with a
relatively high angular resolution. The line has an upper level energy ($E_{\rm up}$) lying $\sim$250 K above the ground and a critical
density ($n_{\rm cr}$) of order $10^6$ cm$^{-3}$, thus is sensitive to hot and dense gas in shocked regions. In addition, the CO (4--3) line
was simultaneously obtained with the dual-frequency receiver, enabling excitation analysis without significant effect from calibration
uncertainties.

\section{Observations} \label{obs}
The observations were conducted on 2010 July 9, with $\sim$0.25 mm precipitable water vapor, using a dual-frequency receiver system on APEX.
The receiver system includes a 1 THz channel, a copy of \emph{Herschel}/HIFI band IV, and a co-aligned 460 GHz channel, designed for pointing
purpose \citep{Leinz10}. Two Fast Fourier Transform Spectrometers, based on the design presented by Klein et al. (2006), were configured to
provide a spectral resolution of 122 kHz ($\sim$0.08 km\,s$^{-1}$) over a 1.6 GHz bandwidth for the 460 GHz channel and a resolution of
183 kHz ($\sim$0.053 km\,s$^{-1}$) over a 2.4 GHz bandwidth for the 1 THz channel. The system temperatures were around 360 K at 460 GHz and
2100 K at 1 THz. The beam efficiencies were determined to be $64\%\pm8\%$ and $21\%\pm7\%$ at 460 GHz and 1 THz, respectively, by observing
Mars and Uranus.

We performed on-the-fly (OTF) mapping observations, covering an area of about $1'\times0.\!'5$. Guided by the outflow elongation
revealed by previous low-$J$ CO observations, the long axis of the mapping area was tilted by $45^{\circ}$ east of north. The effective
integration time on each grid of the OTF map was 13 s. We pointed every hour toward 
(R.A., decl.)$_{\mathrm J2000}=(17^{\mathrm h}20^{\mathrm m}53.\!^{\mathrm s}44, -35^{\circ}46{'}57.\!{''}9$), the peak position of the 
brightest 1.3 mm source \citep[i.e., I-SMA1 in][]{Hunter06} in the region, with the 460 GHz channel in the continuum mode. Furthermore, 
we constructed a pseudo 1 THz continuum map by averaging the line-free channels over an effective bandwidth of $\sim$1.5 GHz, compared that 
pseudo continuum with a 345 GHz continuum map (convolved to the APEX beam of $6.\!''4$) obtained from recent Submillimeter Array 
observations (Q Zhang, priv. comm.), and shifted the final maps by about $3.\!''5$ to match the high-accuracy 345 GHz peak at 
(R.A., decl.)$_{\mathrm J2000}=(17^{\mathrm h}20^{\mathrm m}53.\!^{\mathrm s}34, -35^{\circ}46{'}59.\!{''}1$). Only CO (9--8) emission is 
clearly detected in the THz channel. In the 460 GHz channel, several lines are seen toward the 345 GHz continuum peak, but no line 
contamination is evident over the full observed velocity extent of the CO (4--3) emission. Unless specified, the data are presented in 
$T_{\rm A}^{\ast}$ scales. For a smoothed spectral resolution of 2 km\,s$^{-1}$, the rms noise levels are 0.27 K in CO (9--8) and 0.05 K 
in CO (4--3).

Complementary IRAC 3--8 $\mu$m images were retrieved from the \emph{Spitzer} archive (Program ID: 30154). We use the Post Basic Calibrated
Data products produced by the \emph{Spitzer} Science Center pipeline S18.7.0.

\section{Results} \label{result}
\begin{figure}
\epsscale{.75} \plotone{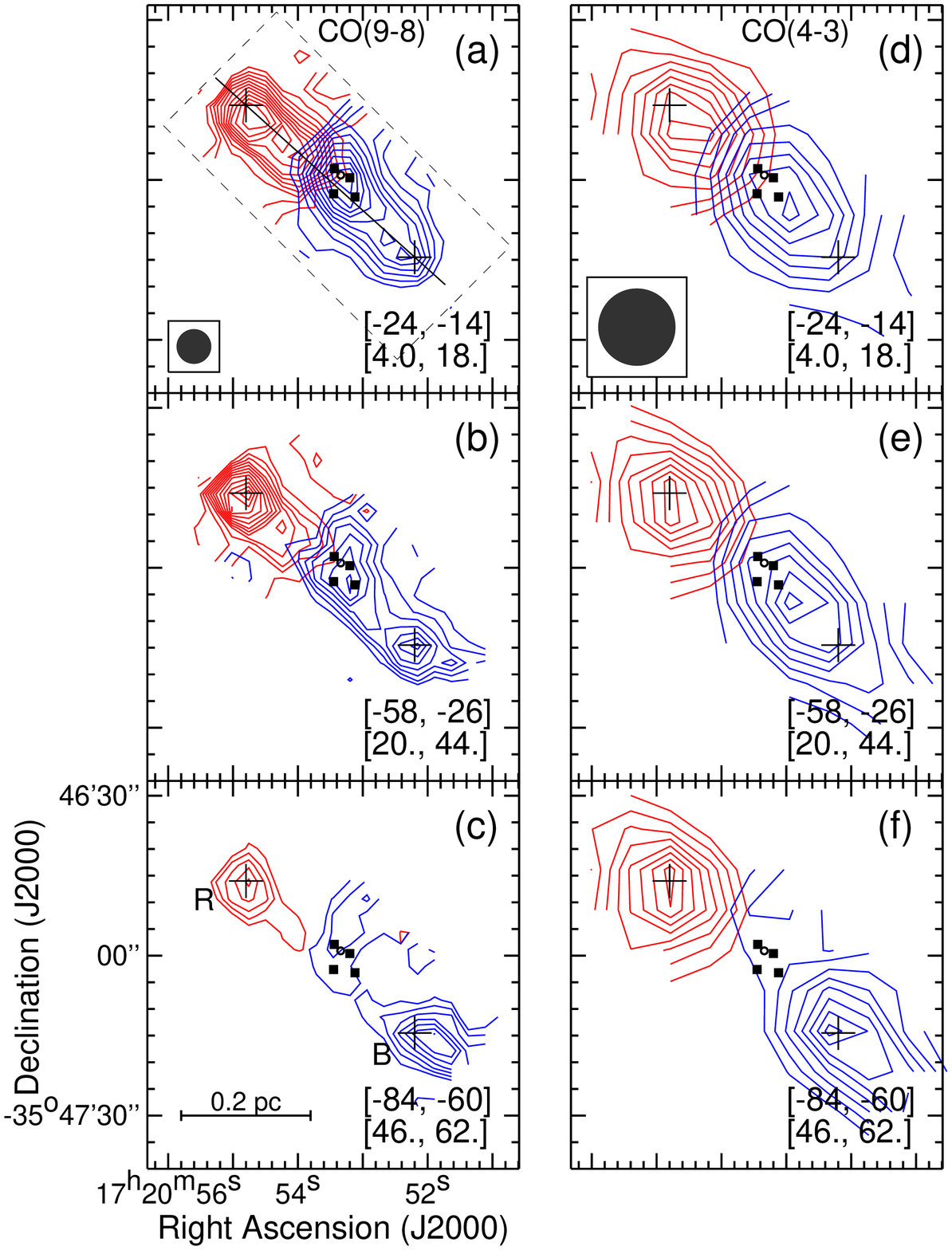} \caption{APEX CO (9--8) emission integrated in three velocity intervals:
(a) s-wing, the lowest contour is 15.6/18.0 K\,km\,s$^{-1}$ (blue/red) and the contour spacing is 3.9/4.5 K\,km\,s$^{-1}$ (blue/red);
(b) f-wing, the lowest contour is 16.5/14.5 K\,km\,s$^{-1}$ (blue/red) and the contour spacing is 6.6/5.8 K\,km\,s$^{-1}$ (blue/red);
(c) EHV, the lowest contour is 11.6/9.6 K\,km\,s$^{-1}$ (blue/red) and the contour spacing is 5.8/4.8 K\,km\,s$^{-1}$ (blue/red).
(d)--(f) The CO (4--3) emission integrated in the same velocity intervals; the equally spaced contour levels go from 25\% to 95\% of the 
peak intensities, which are 173/158 K\,km\,s$^{-1}$ (blue/red) for s-wing, 182/137 K\,km\,s$^{-1}$ (blue/red) for f-wing, and 
76/38 K\,km\,s$^{-1}$ (blue/red) for EHV. Four filled squares denote millimeter continuum sources detected by Hunter et al. (2006). 
A small circle labels the shifted THz continuum peak. Two ``+'' symbols mark approximately the outer peaks of the blue- and redshifted 
lobes, namely ``B'' and ``R'', respectively. A solid line in panel (a) shows the cut of the position-velocity diagram 
(see Figure \ref{pvplot}). 
\label{co98int}}
\end{figure}

The CO (9--8) and (4--3) emission arising from the outflowing gas is detected with very high-velocity line wings, reaching LSR velocities
$-90$ km\,s$^{-1}$ in the blueshifted lobe and $60$ km\,s$^{-1}$ in the redshifted lobe (the ambient cloud velocity is about
$-7.6$ km\,s$^{-1}$). Figure \ref{co98int} shows the emission integrated in three velocity intervals, which are referred as the slow-wing
(s-wing), fast-wing (f-wing), and extremely high-velocity (EHV) components following the terminology of Tafalla et al. (2010). Focusing on
the high-resolution CO (9--8) maps, a bipolar outflow centered around the Trapezium-like millimeter continuum sources is clearly detected.
Provided a pointing uncertainty of 2--3$''$, we refrain from an identification of the outflow driving source from the millimeter sources.
The outflow extends about 0.5 pc from the northeast (NE) to southwest (SW), in an orientation consistent with low- to mid-$J$ CO
observations. A blueshifted structure extending from the millimeter sources to the northwest also shows EHV emission, and is likely
attributed to another outflow. Since the feature extends beyond our mapping area, it is not further analyzed in this work.

\begin{figure}
\epsscale{1} \plottwo{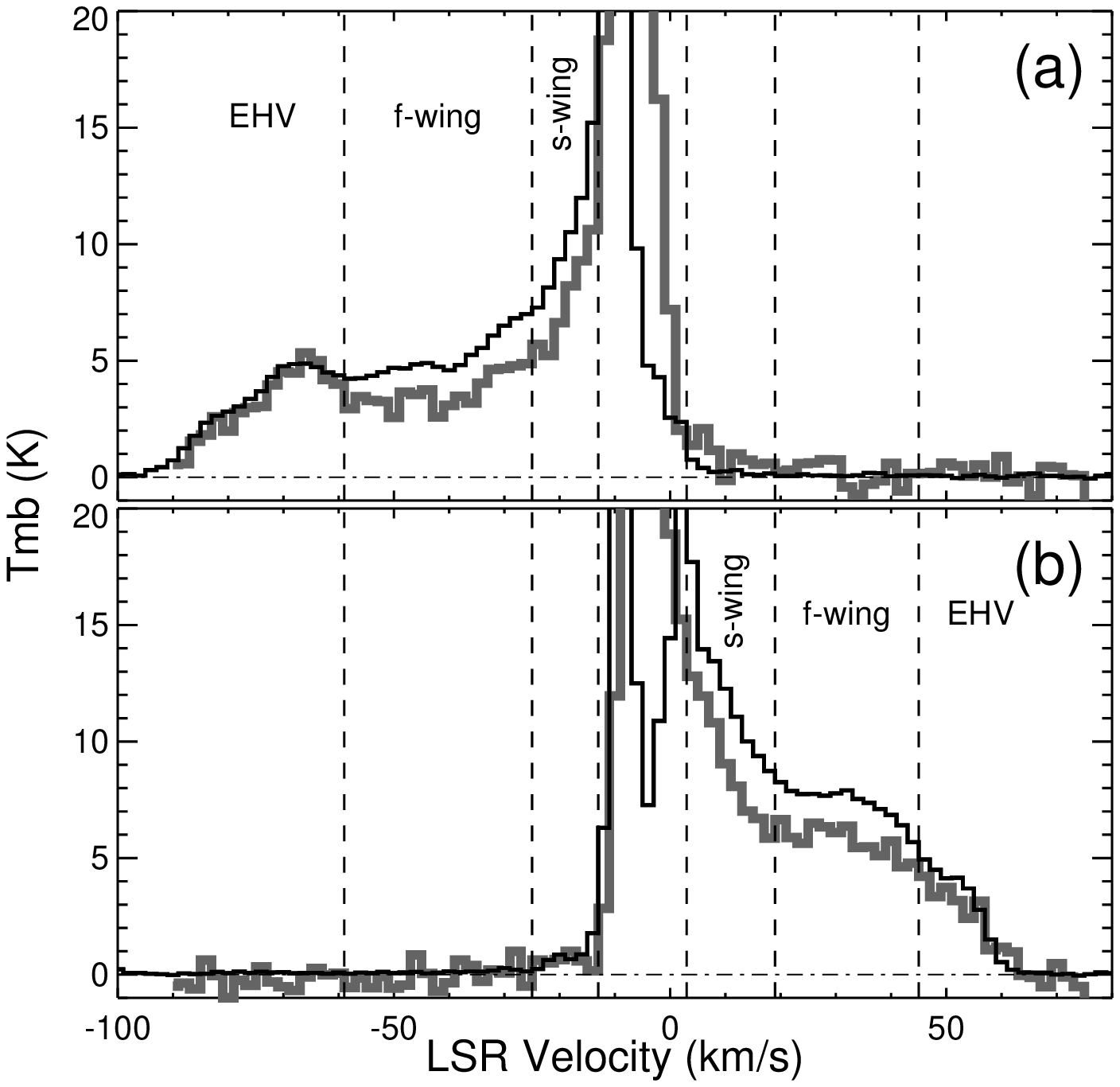}{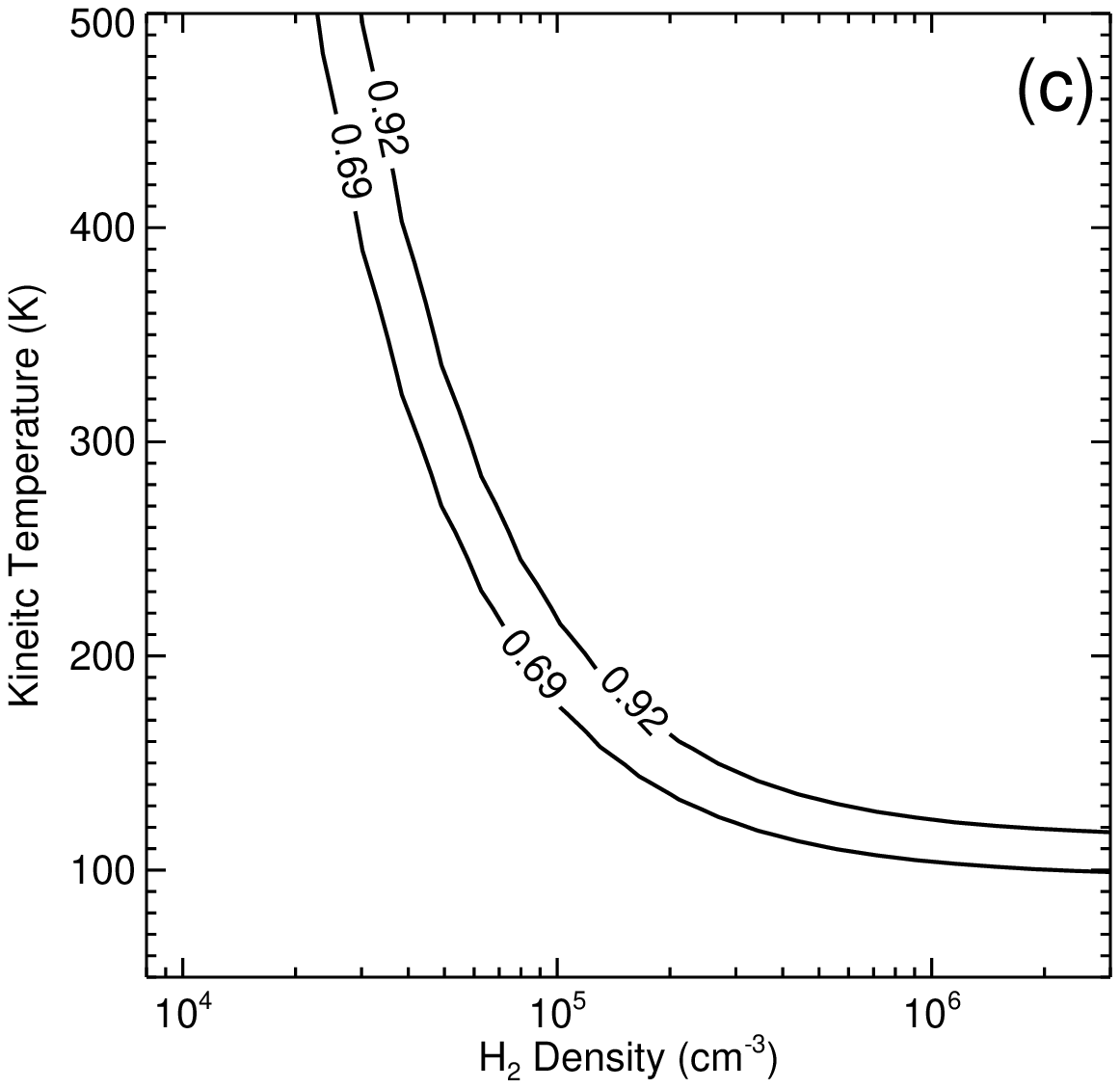} \caption{(a) The CO (9--8) and (4--3) spectra extracted from the ``B'' position (Figure \ref{co98int})
shown in gray and black histograms, respectively; the CO (9--8) data have been convolved to the CO (4--3) beam, which is $14.\!''4$ at FWHM;
the vertical dashed lines delineate the three velocity intervals (see Figure \ref{co98int} and Section \ref{result}). (b) Same as (a),
but extracted from the ``R'' position (Figure \ref{co98int}). (c) Curves of constant CO (9--8)/(4--3) ratios as a function of H$_2$ density
and kinetic temperature, for an adopted CO column density to line width ratio of $10^{16}$ cm$^{-2}$/(km\,s$^{-1}$).
\label{spectra}}
\end{figure}

Figures \ref{spectra}(a), \ref{spectra}(b) show the CO (9--8) and (4--3) spectra extracted toward the outer peak on each of the blue- and
redshifted lobes (marked as ``B'' and ``R'' in Figure \ref{co98int}). The two transitions have similar profiles: the s-wing component has its
brightness fast decreasing as the velocity increases; the f-wing component appears as a plateau with relatively flat brightness variation;
the EHV component has greater CO (9--8)/(4--3) ratios, and shows a secondary peak in the blueshifted lobe. The CO (9--8)/(4--3) ratios
averaged within each of the three velocity intervals are measured to be 0.70 (s-wing), 0.69 (f-wing), 0.92 (EHV) toward ``B'' and
0.79 (s-wing), 0.81 (f-wing), 0.92 (EHV) toward ``R''. Again, greater ratios are found in the EHV component, which imply a higher excitation
condition for the EHV gas. We performed large-velocity-gradient (LVG) calculations on the line ratios using the RADEX code \citep{vdTak07}.
Figure \ref{spectra}(c) shows the CO (9--8)/(4--3) ratios as a function of gas density and kinetic temperature. The adopted CO column density
to line width ratio is $10^{16}$ cm$^{-2}$/(km\,s$^{-1}$), corresponding to a total column density on the order of $10^{17}$ cm$^{-2}$ as
constrained by Leurini et al. (2006) based on $^{13}$CO observations. The calculations show that for the observed range of line ratios,
the gas density is $>10^4$ cm$^{-3}$ and the temperature is $\gtrsim100$ K. To further constrain the physical conditions of the outflowing gas,
it is instructive to adopt a typical CO abundance of $10^{-4}$ and assume that the gas dimension along the light-of-sight is similar to that
in the plane-of-sky. In this situation the CO column density and H$_2$ density are connected by a size scale, and that scale is used to
estimate the beam filling factor. We then ran a series of LVG models exploring the parameter space of kinetic temperature, H$_2$ density,
and CO column density, and compared the brightness temperatures predicted by the models with the corrected antenna temperatures ($T_{\rm mb}$
divided by a beam filling factor). We find that the models with a density of $\sim$2--$3\times10^4$ cm$^{-3}$ and a kinetic temperature of
$\sim$400 K appear to reasonably agree with the observed EHV emission, while the models with a similar density but a lower temperature of
$\sim$300 K are able to reproduce the s-wing/f-wing emission.

\begin{figure}
\epsscale{1} \plotone{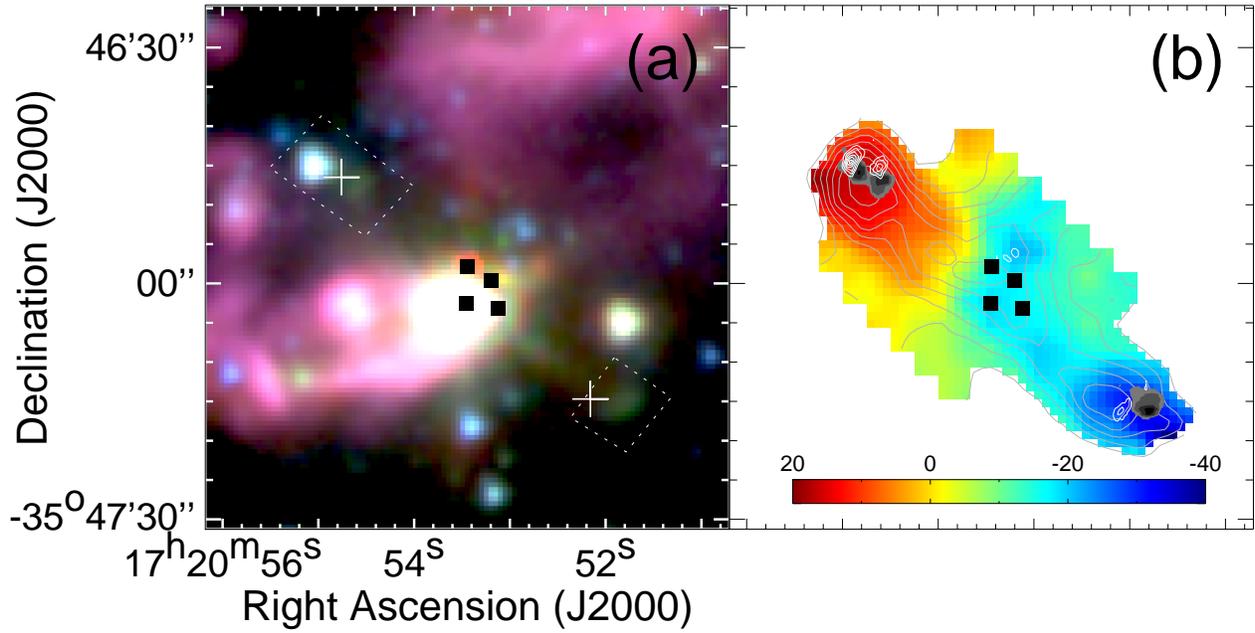} \caption{(a) \emph{Spitzer} three-color composite image with the 3.6, 4.5, and 5.8 $\mu$m emission coded in
blue, green, and red, respectively; dotted lines outline two areas where the 4.5 $\mu$m/3.6 $\mu$m band ratio is computed; other symbols are
the same as shown in Figure \ref{co98int}. (b) The colored image shows the 1st moment map of the CO (9--8) emission, and the gray contours
show the 0th moment map with the lowest contour of 110 K\,km\,s$^{-1}$ and the contour spacing of 22 K\,km\,s$^{-1}$; the gray scale represent
the 4.5 $\mu$m/3.6 $\mu$m band ratio at levels of 1.7, 1.85, 2.0, and 2.15; white contours represent the NH$_3$ (3,3) maser emission
originally presented in Beuther et al. (2007).
\label{spitzer}}
\end{figure}

Figure \ref{spitzer} shows a comparison between the CO (9--8) emission and shock activities probed by mid-IR and NH$_3$ maser observations.
In Figure \ref{spitzer}(a), the \emph{Spitzer} three-color composite image reveals a pair of bow-shaped tips around the CO (9--8) peaks in
excess 4.5 $\mu$m (\emph{green}) emission, which often traces jets and bow-shocks in protostellar outflows
\citep[e.g.,][]{Noriega04,Smith06,Qiu08}. Several authors have detected 2.12 $\mu$m H$_2$ knots in this region
\citep{Davis95,Persi96,Eisloffel00,Seifahrt08}. Two bow-shaped H$_2$ knots along the CO outflow axis
\citep[denoted ``$\alpha$'' and ``$\beta$'' in][]{Seifahrt08} coincide with the 4.5 $\mu$m emission knots reported here. This suggests that
the excess 4.5 $\mu$m emission can be mostly due to shocked H$_2$ lines \citep{Smith05,Smith06,DeBuizer10}, but the contribution from
CO $\nu$=1--0 or H{\scriptsize I} Br$\alpha$ lines cannot be ruled out without spectroscopic observations.
Furthermore, Figure \ref{spitzer}(b) shows the 4.5 $\mu$m to 3.6 $\mu$m emission ratio to highlight the
shocked emission \citep{Teixeira08,Takami10}. While field stars have the brighter emission falling in the 3.6 $\mu$m band, shocked emission
appears most prominent in the 4.5 $\mu$m band, thus Figure \ref{spitzer}(b) not only shows the bow-shaped tips seen in Figure \ref{spitzer}(a)
but also unveils a new feature which is suppressed in Figure \ref{spitzer}(a) by a nearby bright star.
Each of the three 4.5 $\mu$m excess features is associated with an NH$_3$ (3,3) maser spot \citep{Beuther07}.
Compared to the 1st moment (intensity-weighted velocity) map of the CO (9--8) emission, the shocked features seen in IR and maser emission
coincide with the outer parts of the outflow lobes where most of the gas moves at high velocities.

\section{Discussion} \label{dis}
 The most surprising result of our observations is the detection of high-velocity CO (9--8) emission (f-wing and EHV) over a spatial extent
of order 0.5 pc. This provides direct evidence for the presence of a highly-excited component in a bipolar outflow. The dynamical timescale
of the CO (9--8) outflow, derived from the flow radius of $\sim$0.22 pc divided by the outflow terminal velocity of $\sim$77 km\,s$^{-1}$
\citep{Lada85}, is about $2.6\times10^3$ yr. This agrees with previous estimates of 2--$4\times10^3$ yr based on the CO (2--1), (3--2), and
(4--3) observations \citep{Bachiller90,McCutcheon00,Leurini06}. Therefore the outflow is very young, preferring an evolutionary stage prior to
the formation of an UC H{\scriptsize II} region for its driving source. Compared to the mid-IR and NH$_3$ maser observations, the outflow
seems to be closely related to bow shocks. The CO (9--8) observations have a $\sim6''$ resolution, allowing us to further explore the
kinematics of the outflow and its driving mechanism.

\begin{figure}
\epsscale{1} \plotone{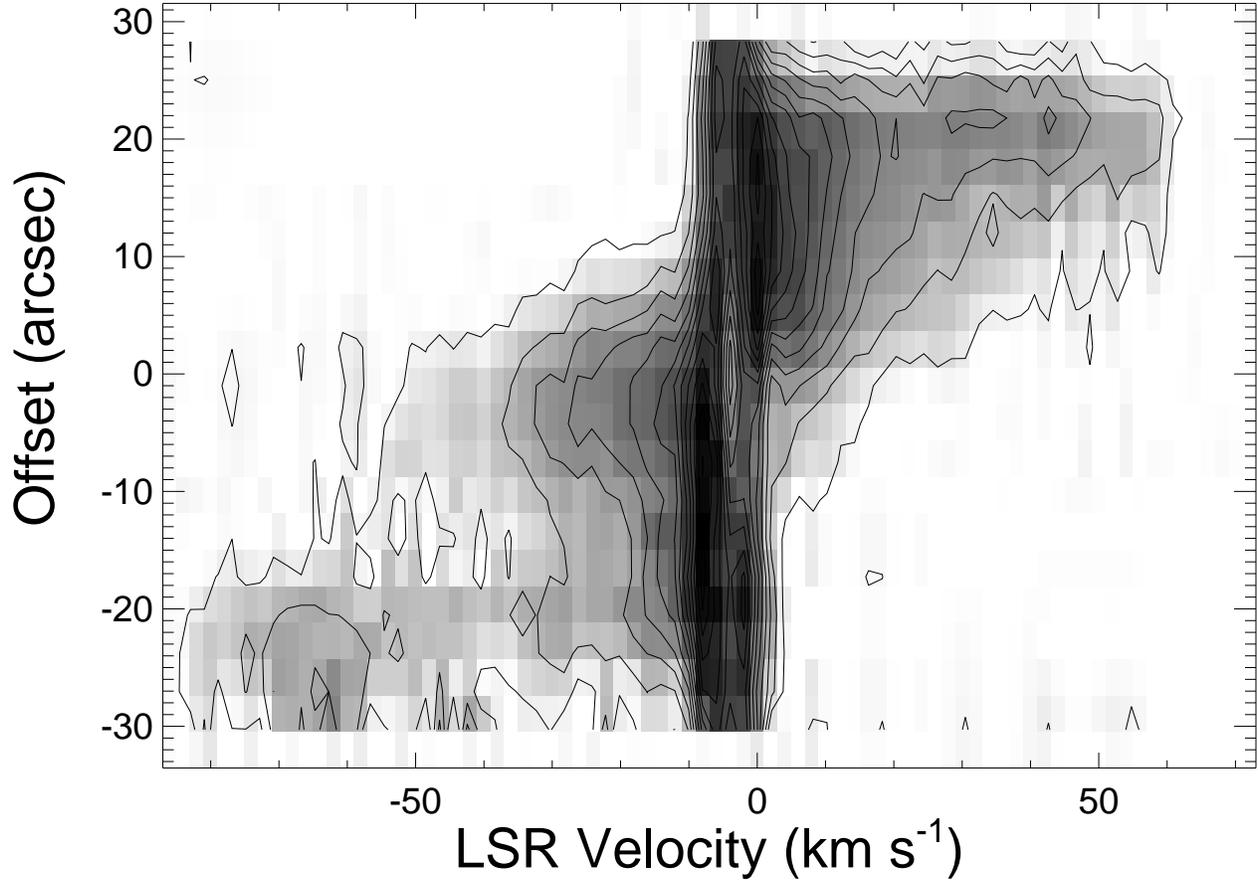} \caption{CO (9--8) position-velocity diagram constructed along the direction going through the peaks of the
outflow lobes (see Figure \ref{co98int}(a)); the lowest contour and the contour spacing are 0.75 K; the offsets are measured with
respective to the midpoint of the two peaks, and should roughly represent the distances to the outflow central source.
\label{pvplot}}
\end{figure}

Figure \ref{pvplot} shows the position-velocity (P-V) diagram of the CO (9--8) emission, constructed approximately along the outflow axis.
Overall, the outflow velocities (relative to the ambient cloud velocity) are greater at positions farther away from the central source. Such
a velocity structure is also discernable in the 1st-moment map shown in Figure \ref{spitzer}(b). In particular, the maximum velocities in the
redshifted lobe roughly linearly increase with the distances from the central source, following a ``Hubble'' law. Theoretically, outflows
forming from jet-driven bow shocks have been shown to exhibit Hubble-like P-V relations, where the forward velocities of the gas around the
bow's head are greater than along the bow wings \citep{Zhang97,Smith97,Downes99,Lee01}. On the other hand, Stahler (1994) obtained the Hubble
P-V relation in a model of ambient gas entrained by a turbulent jet. However, as pointed out by Lada \& Fich (1996), a critical assumption of
Stahler's calculation is that the gas mass in every slab perpendicular to the jet axis has the same power-law distribution with velocity,
which is not seen in our CO (9--8) or (4--3) data. The P-V pattern we observe is very similar to that reproduced by the bow-shock calculations
of Zhang \& Zheng (1997) and simulations of Smith et al. (1997) and Downes \& Ray (1999). We also note that the P-V pattern shown in
Figure \ref{pvplot} closely resembles that of the NGC 2264G outflow, which is a Class 0 outflow driven by a precessing jet
\citep{Lada96,Teixeira08}. Moreover, the bow-shaped tips seen in the IR independently suggest the association between bow shocks and the
CO (9--8) emission. Thus, we suggest that the gas probed by the CO (9--8) line wings is accelerated by bow shocks created by a jet impacting
the ambient material. The nature of the EHV gas, however, is less clear. From Figure \ref{spectra}, the terminal velocities of the EHV
component reach $>100$ km\,s$^{-1}$ if the projection effect is taken into account. Non-dissociative C-type shocks are expected to accelerate
ambient gas to velocities up to $\sim50$ km\,s$^{-1}$. A recent chemical investigation of EHV low-mass outflows challenges the
hypothesis that molecules in EHV gas reform behind dissociative J-type shocks \citep{Tafalla10}. It is likely that the EHV gas has an origin
close to the central source rather than being accelerated ambient gas \citep{Qiu09,Tafalla10}.

\begin{figure}
\epsscale{1} \plotone{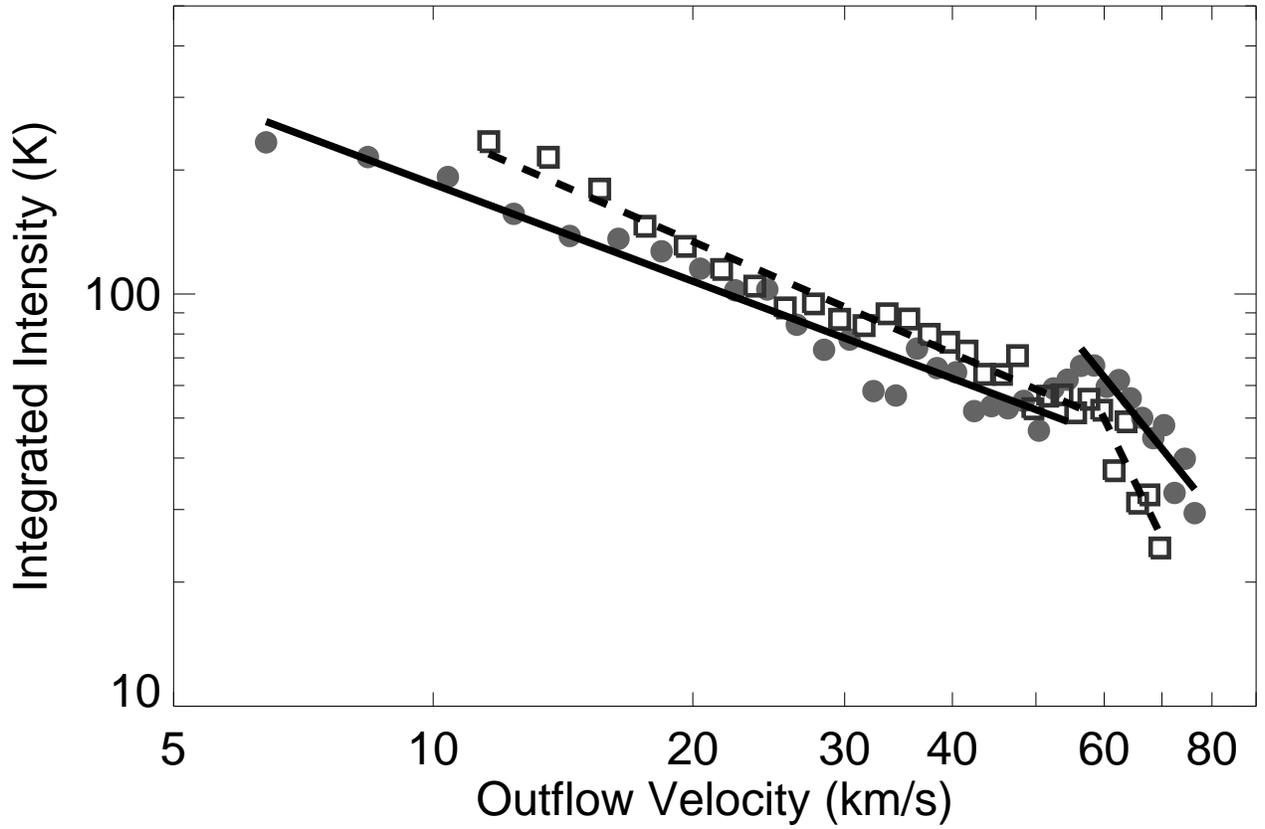} \caption{Spatially integrated intensity of the CO (9--8) emission as a function of outflow velocity.
Filled circles and open squares denote the blue- and redshifted emission measured within each 2 km\,s$^{-1}$ velocity channel, respectively.
Power-law fittings to the data are overlaid in solid (blue) and dashed (red) lines. \label{mvplot}}
\end{figure}

Most low-$J$ CO observations of molecular outflows show power-law intensity-velocity (I-V) relations, $I_{\rm CO}(v){\propto}v^{-\gamma}$,
with the power-law index, $\gamma$, typically around 2 and in some cases steepening to larger values at velocities higher than
10--30 km\,s$^{-1}$ \citep[e.g.,][]{Stahler94,Lada96,Davis98,Ridge01}. If the line wings are optically thin and the CO abundance
and excitation temperature do not vary much, as assumed in many studies, the shape of I-V plots represents mass-velocity
distributions \citep[see, e.g., a compilation by][]{Richer00}. The I-V property provides an important test for models of molecular outflow
acceleration \citep[e.g.,][]{Downes03,Keegan05}. In Figure \ref{mvplot}, we plot the spatially integrated CO (9--8) intensity as
a function of velocity. For $v\lesssim55$ km\,s$^{-1}$, the intensity distribution can be described by power-laws with $\gamma=0.78\pm0.04$ and
$0.90\pm0.04$ for the blue- and redshifted lobes, respectively. At higher velocities, the intensity rapidly drops, and the power-law fits give
$\gamma=2.59\pm0.33$ (blue) and $4.32\pm1.27$ (red). The I-V relation in CO (4--3) is similar, with $\gamma=1$ in each lobe up to
$\sim55$ km\,s$^{-1}$ and steepening further out. In addition, Ridge \& Moore (2001) reported $\gamma\approx1$ for $v<50$ km\,s$^{-1}$ with
the CO (2--1) observations of this outflow. Thus, the I-V relation of the NGC 6334 I outflow, measured in low- to high-$J$ CO lines for
velocities up to $\sim55$ km\,s$^{-1}$, appears to be among the shallowest known and is shallower than most theoretical predictions
\citep[see a recent review by][]{Arce07}. Simulations of collimated outflows driven by dense jets predict $\gamma$ close to 1
\citep[jet to ambient gas density ratio set to 10,][]{Rosen04}. But the thrust of the simulated jets can only drive low-mass outflows, and
the $\gamma\sim1$ power-law in the simulations occurs at velocities up to 10 km\,s$^{-1}$, far lower than the value observed here
(55 km\,s$^{-1}$). It is therefore yet to be confirmed whether the observed shallowness in the I-V relation can be a consequence of the high
density of an underlying jet. Finally, $I_{\rm CO}(v)$ at $\gtrsim55$ km\,s$^{-1}$ behaves dramatically different from that at lower velocities.
It probably reflects a different nature of the EHV gas as discussed above.

\section{Summary} \label{conclu}
The observations reveal a jet-like and extremely high-velocity outflow originating from a high-mass protostar. The CO (9--8) map presented
here is the very first THz frequency map of a molecular outflow. It indicates that a highly-excited CO line
($E_{\rm up}\sim250$ K, $n_{\rm cr}\sim10^6$ cm$^{-3}$) can still have broad line wings tracing parsec scale outflows. LVG calculations of the
CO (9--8)/(4--3) line ratios give kinetic temperatures $>100$ K and H$_2$ densities $>10^4$ cm$^{-3}$. The spatial-velocity structure of this
relatively hot and dense outflowing gas suggests that the outflow is driven by jet bow-shocks. The detection of IR bow-shaped tips coincident
with the CO outflow lobes further supports this scenario. It is worth noting that the outflow shows an exceptionally shallow I-V relation in
low- to high-$J$ CO lines, though whether that shallowness is caused by a high density of the jet is to be addressed. The outflow also
possesses EHV gas. The CO (9--8)/(4--3) ratios and the I-V relation of the EHV emission both appear to be different from lower velocity line
wings, which may suggest a different origin (close to the central protostar) of the EHV gas.

Finally, the observations demonstrate that under excellent weather conditions, high-quality THz observations can be performed with the new THz
receiver on APEX.

\acknowledgments This work is based in part on observations made with the Spitzer Space Telescope, which is operated by the Jet Propulsion
Laboratory, California Institute of Technology under a contract with the National Aeronautics and Space Administration (NASA).

\clearpage

\begin{thebibliography}{}
\bibitem[Arce et al. 2007]{Arce07}Arce, H. G., Shepherd, D. S., Gueth, F., Lee, C.-F., Bachiller, R., Rosen, A., \& Beuther, H. 2007, in Protostars and Planets V., ed. B. Reipurth, D. Jewitt, \& K. Keil (Tucson, AZ: Univ. Arizona Press), 245
\bibitem[Bachiller \& Cernicharo 1990]{Bachiller90}Bachiller, R. \& Cernicharo, J. 1990, \aap, 239, 276
\bibitem[Beuther et al. 2007]{Beuther07}Beuther, H., Walsh, A. J., Thorwirth, S., Zhang, Q., Hunter, T. R., Megeath, S. T., \& Menten, K. M. 2007, \aap, 466, 989
\bibitem[Beuther et al. 2008]{Beuther08}Beuther, H., Walsh, A. J., Thorwirth, S., Zhang, Q., Hunter, T. R., Megeath, S. T., \& Menten, K. M. 2008, \aap, 481, 169
\bibitem[Davis \& Eisl\"{o}ffel 1995]{Davis95}Davis, C. J. \& Eisl\"{o}ffel, J. 1995, \aap, 300, 851
\bibitem[Davis et al. 1998]{Davis98}Davis, C. J., Moriarty-Schieven, G., Eisl\"{o}ffel, J., Hoare, M. G., \& Ray, T. P. 1998, \apj, 115, 1118
\bibitem[De Buizer \& Vacca 2010]{DeBuizer10}De Buizer, J. M. \& Vacca, W. D. 2010, \aj, 140, 196
\bibitem[Downes \& Cabrit 2003]{Downes03}Downes, T. P. \& Cabrit, S. 2003, \aap, 403, 135
\bibitem[Downes \& Ray 1999]{Downes99}Downes, T. P. \& Ray, T. P. 1999, \aap, 345, 977
\bibitem[Eisl\"{o}ffel et al. 2000]{Eisloffel00}Eisl\"{o}ffel, J., Smith, M. D., \& Davis, C. J. 2000, \aap, 359, 1147
\bibitem[Emprechtinger et al. 2010]{Emprechtinger10}Emprechtinger, M. et al. 2010, \aap, 521, L28
\bibitem[Giannini et al. 2001]{Giannini01}Giannini, T., Nisini, B., \& Lorenzetti, D. 2001, \apj, 555, 40
\bibitem[Hunter et al. 2006]{Hunter06}Hunter, T. R., Brogan, C. L., Megeath, S. T., Menten, K. M., Beuther, H., \& Thorwirth, S. 2006, \apj, 649, 888
\bibitem[Kawamura et al. 2002]{Kawamura02}Kawamura, J. et al. 2002, \aap, 394, 271
\bibitem[Keegan \& Downes 2005]{Keegan05}Keegan, R. \& Downes, T. P. 2005, \aap, 437, 517
\bibitem[Klein et al. 2006]{Klein06}Klein, B., Philipp, S. D., Kr{\"a}mer, I., Kasemann, C., G{\"u}sten, R., \& Menten, K. M. 2006, L29
\bibitem[Lada 1985]{Lada85}Lada, C. J. 1985, \araa, 23, 267
\bibitem[Lada \& Fich 1996]{Lada96}Lada, C. J. \& Fich, M. 1996, \apj, 459, 638
\bibitem[Lee et al. 2001]{Lee01}Lee, C.-F., Stone, J. M., Ostriker, E. C., \& Mundy, L. G. 2001, \apj, 557, 429
\bibitem[Leinz et al. 2010]{Leinz10}Leinz, C. et al. 2010, in Proceedings of the 21st International Symposium on Space Terahertz Technology (http://www.physics.ox.ac.uk/stt2010/proceedings.aspx), p.130-135
\bibitem[Leurini et al. 2006]{Leurini06}Leurini, S., Schilke, P., Parise, B., Wyrowski, F., G\"{u}sten, R., \& Philipp, S. 2006, \aap, 454, L83
\bibitem[Marrone et al. 2004]{Marrone04}Marrone, D. P. et al. 2004, \apj, 612, 940
\bibitem[McBreen et al. 1979]{McBreen79}McBreen, B., Fazio, G. G., Stier, M., \& Wright, E. L. 1979, \apj, 232, L183
\bibitem[McCutcheon et al. 2000]{McCutcheon00}McCutcheon, W. H., Sandell, G. Matthews, H. E., Kuiper, T. B. H., Sutton, E. C., Danchi, W. C., \& Sato, T. 2000, \aap, 316, 152
\bibitem[Neckel 1978]{Neckel78}Neckel, T. 1978, \aap, 69, 51
\bibitem[Noriega-Crespo et al. 2004]{Noriega04}Noriega-Crespo, A. et al. 2004, \apjs, 154, 352
\bibitem[Persi et al. 1996]{Persi96}Persi, P., Roth, M., Tapia, M., Marenzi, A. R., Felli, M., Testi, L, \& Ferrari-Toniolo, M. 1996, \aap, 307, 591
\bibitem[Qiu et al. 2008]{Qiu08}Qiu, K. et al. 2008, \apj, 685, 1005
\bibitem[Qiu \& Zhang 2009]{Qiu09}Qiu, K. \& Zhang, Q. 2009, \apj, 702, L66
\bibitem[Richer et al. 2000]{Richer00}Richer, J., Shepherd, D., Cabrit, S., Bachiller, R., \& Churchwell, E. 2000, in Protostars and Planets IV, ed. V. Mannings, A. Boss, \& S. Russell (Tucson, AZ: Univ. Arizona Press), 867
\bibitem[Ridge \& Moore 2001]{Ridge01}Ridge, N. A. \& Moore, T. J. T. 2001, \aap, 378, 495
\bibitem[Rodr\'{\i}guez et al. 1982]{Rodriguez82}Rodr\'{\i}guez, L. F., Cant\'{o}, J., \& Moran, J. M. 1982, \apj, 255, 103
\bibitem[Rosen \& Smith 2004]{Rosen04}Rosen, A. \& Smith, M. D. 2004, \aap, 413, 593
\bibitem[Seifahrt et al. 2008]{Seifahrt08}Seifahrt, A. et al. 2008, J. Phys. Conf. Ser., 131, 012030
\bibitem[Smith et al. 2006]{Smith06}Smith, H. A., Hora, J. L., \& Marengo, M. 2006, \apj, 645, 1264
\bibitem[Smith \& Rosen 2005]{Smith05}Smith, M. D. \& Rosen, A. 2005, \mnras, 357, 1370
\bibitem[Smith et al. 1997]{Smith97}Smith, M. D., Suttner, G., \& Yorke, H. W. 1997, \aap, 323, 223
\bibitem[Stahler 1994]{Stahler94}Stahler, S. W. 1994 \apj, 422, 616
\bibitem[Takami et al. 2010]{Takami10}Takami, M., Karr, J. L., Koh, H., Chen, H.-H., \& Lee, H.-T. 2010, \apj, 720, 155
\bibitem[Tafalla et al. 2010]{Tafalla10}Tafalla, M., Santiago-Garc\'{\i}a, J., Hacar, A., \& Bachiller, R. 2010, \aap, 522, 91
\bibitem[Teixeira et al. 2008]{Teixeira08}Teixeira, P. S., Coey, C. M., Fich, M., \& Lada, C. J. 2008, \mnras, 384, 71
\bibitem[van der Tak et al. 2007]{vdTak07}van der Tak, F. F. S., Black, J. H., Sch\"{o}ier, F. L., Jansen, D. J., \& van Dishoeck, E. F. 2007, \aap, 468, 627
\bibitem[van Kempen et al. 2010]{vKempen10}van Kempen, T. A. et al. 2010, \aap, 518, L121
\bibitem[van der Wiel et al. 2010]{vdWiel10}van der Wiel, M. H. D. et al. 2010, \aap, 521, L43
\bibitem[Wiedner et al. 2006]{Wiedner06}Wiedner, M. C. et al. 2006, \aap, 454, L33
\bibitem[Yildiz et al. 2010]{Yildiz10}Yildiz, U. A. et al. 2010, \aap, 521, L40
\bibitem[Zhang \& Zheng 1997]{Zhang97}Zhang, Q. \& Zheng, X. 1997, \apj, 474, 719
\end{thebibliography}
\end{document}